\documentclass[preprint,showpacs,preprintnumbers,amsmath,amssymb]{revtex4}

\usepackage{graphicx}
\usepackage{epsfig}		
\usepackage{dcolumn}
\usepackage{bm}

\def\V{\mbox{\tiny $V$}}
\def\U{\mbox{\tiny $U$}}
\def\B{\mbox{\tiny $B$}}
\def\R{\mbox{\tiny $R$}}
\def\L{\mbox{\tiny $L$}}

\begin{document}

\title{OBTAINING THE GAUGE INVARIANT KINETIC TERM FOR A $SU(n)_U \otimes SU(m)_V$ LAGRANGIAN}

\author{A. E. Bernardini}
\affiliation{Instituto de F\'{\i}sica Gleb Wataghin, UNICAMP,\\
PO Box 6165, 13083-970, Campinas, SP, Brazil.}
\email{alexeb@ifi.unicamp.br}

\date{\today}

\begin{abstract}
We propose a generalized way to formally obtain the gauge invariance of the kinetic part of a field Lagrangian over which a gauge transformation ruled by an $SU(n)_{U} \otimes SU(m)_{V}$ coupling symmetry is applied.  
As an illustrative example, we employ such a formal construction for reproducing the standard model Lagrangian.
This generalized formulation is supposed to contribute for initiating the study of gauge transformation applied to generalized $SU(n)_{U} \otimes SU(m)_{V}$ symmetries as well as for complementing an introductory study of the standard model of elementary particles.
\end{abstract}

\maketitle

Physicists mostly agree that conservation laws of Nature enjoy a high degree of symmetry, that is, the formulation of these laws is unchanged when various transformations are performed.
Presence of symmetry implies absence of complicated and irrelevant structure, and our conviction that this is fundamentally true reflects the belief that Nature in its fundamental workings is essentially simple.
However, we must also recognize that observed physical phenomena rarely exhibit overwhelming regularity.
Therefore, at the very same time that we construct a physical theory with an intrinsic symmetry, we must find a way to break the symmetry in physical consequences of the model.

The relations between symmetries and conservation laws are more conveniently verified in the Hamiltonian formalism: a conservation law for a physical system is directly related to the invariance of the system Hamiltonian $H$ under a set of transformations.
Formally, for a set of transformations represented by an operator $S$, the invariance is expressed by
\begin{equation}
SHS^{\dag}=H.
\label{sim1} 
\end{equation}
In classical physics, the principal mechanism for symmetry breaking is through boundary and initial conditions on dynamical equations of motion.
For example, Newton's rotationally symmetric gravitational equations admit the rotationally non-symmetric solutions that describe actual orbits in the solar system, when appropriate,
rotationally non-symmetric, initial conditions are posited.

The construction of physically successful quantum field theories makes use of symmetry for yet another reason.
Quantum field theory models are notoriously difficult to solve and also explicit calculations are beset by infinities.
Thus far we have been able to overcome these two obstacles only when the models possess a high degree of symmetry,
which allows unraveling the complicated dynamics and taming the infinities by renormalization.
Our present-day model for quarks, leptons, and their interactions exemplifies this by enjoying a variety of chiral,
scale/conformal, and gauge symmetries.
But to agree with experiments, most of these symmetries must be absent in the solutions.
At present we have available two mechanisms for achieving this necessary result.
One is spontaneous symmetry breaking, which relies on energy differences between symmetric and non-symmetric solutions:
the dynamics may be such that the non-symmetric solution has lower energy than the symmetric one,
and the non-symmetric one is realized in Nature while the symmetric solution is unstable. 
At a more fundamental level, when the Lagrangian density $\mathcal{L}$ of the system contains
a term which is not invariant under the relevant transformation or, when the
invariance condition is satisfied by $\mathcal{L}$ but the physical vacuum does not remain invariant.
In the first case, the symmetry is explicitly broken, in the second one, we say it is spontaneously 
broken\footnote{Formally, the correct term is {\em realized} and not {\em broken}.} \cite{Pes95}.
The second mechanism is anomalous or quantum mechanical symmetry breaking,
which uses the infinities of quantum theory to effect a violation of the correspondence principle:
the symmetries that appear in the model before quantization disappear after quantization,
because the renormalization procedure (needed to tame the infinities and well define the theory) cannot be carried out
in a fashion that preserves the symmetries.

Since we can find several forms and formulations for symmetries,
a general classification should be implemented.
By taking into account the parameters through which a symmetry is defined, 
we can have discrete or continuous symmetries.
Some typical examples of discrete symmetries are the parity $P$, the charge conjugation $C$
and the time-reversal $T$.
As an example of continuous symmetry, we consider the rotation operations
generically written as $R(\theta)$, where the $\theta$ angle can be varied continuously.
Among continuous symmetries we can identify the external (or geometric) ones and
the internal ones.
The external symmetries act on the space-time, i. e. they refer to the invariance of the system
by some space-time coordinate transformation. 
The Lorentz transformation is a current example.
Otherwise, the internal symmetries act on a set of internal quantum numbers, for example,
the electric charge quantum number.
The symmetries can still be subdivided into other two different groups:
the global symmetries, where the transformation parameters do not depend on the space-time coordinates
$x_{\mu}$, and the local symmetries, where the transformations are not equivalent for different
space-time coordinates since the transformation parameters depend on $x_{\mu}$.
An important physical aspect concern with the possibility of accommodating 
the known particle spectrum in irreducible representations of an internal symmetry group.
In particular, the called {\em gauge} symmetries are often used as a dynamic principle 
to build a field theory.
The idea that fundamental physical interactions are determined by a gauge symmetry has achieved 
an essential importance along the years.
The theories which present some gauge symmetry has the important property of
reducing considerably the possible forms of coupling among the particles in the theory.
Moreover, it introduces the unification character among different classes of interactions since they
thus present an equivalent geometric structure.

In this interim, we intend to elaborate on the construction of the kinetic part of a Lagrangian by considering the
invariance under unitary transformations which characterize the symmetries described by a generic $SU(n)_{U} \otimes SU(m)_{V}$ symmetry.
We do not concern with interaction terms since we are just interested in introducing a useful tool with a simplified notation
for the construction of a formalism already known.
To finalize, as an instructive example, we consider
the $SU(2)_{L} \otimes U(1)_{Y}$ GWS theory for electroweak interactions as a particular application 
of our generalized discussion.

First of all, we shall succinctly illustrate
the construction of the Lagrangian kinetic term which is invariant under
transformations ruled by a generic $SU(n)_{U} \otimes SU(m)_{V}$ coupling symmetry.
By considering a fermionic field $\Psi$ as a fundamental representation of
$SU(n)_{U} \otimes SU(m)_{V}$ where a
global gauge transformation is represented by
\begin{equation}
\Psi \rightarrow \Psi^{\prime} = U \Psi V,
\label{aabaa}
\end{equation}
with $U$ and $V$ respectively describing distinct commuting symmetry operations (unitary rotation matrix)
of $SU(n)_{U}$ and  $SU(m)_{V}$,
we can write a $SU(n)_{U} \otimes SU(m)_{V}$ gauge global invariant Lagrangian term
as
\begin{eqnarray}
\mathcal{L}_1 &=& i\, Tr [\overline{\Psi}^{\prime}(\gamma^\mu \partial_\mu)\Psi^{\prime}]
=i\, Tr [V^{\dagger}\overline{\Psi}U^{\dagger}(\gamma^\mu \partial_\mu)U \Psi V]
=i\, Tr [V^{\dagger}\overline{\Psi}(\gamma^\mu \partial_\mu)\Psi V]
=i\, Tr [\overline{\Psi}(\gamma^\mu \partial_\mu)\Psi].
\label{avfr}
\end{eqnarray}
By assuming a $SU(n)_{U} \otimes SU(m)_{V}$ local gauge invariance,
the same element of (\ref{avfr}) must transform as
\begin{eqnarray}
\mathcal{L}_1 &=& i\, Tr [\overline{\Psi}^{\prime}(\gamma^\mu D^{\prime}_\mu)\Psi^{\prime}]
=i\, Tr [V^{\dagger}\overline{\Psi}U^{\dagger}U (\gamma^\mu D_\mu) \Psi V]
=i\, Tr [V^{\dagger}\overline{\Psi} (\gamma^\mu D_\mu) \Psi V]
\nonumber \\  &=&  i\, Tr [\overline{\Psi} (\gamma^\mu D_\mu) \Psi]~~~~~~~~\Rightarrow~ D^{\prime}_\mu \Psi^{\prime} = U D_\mu \Psi V.
\label{defdef}
\end{eqnarray}
where the correlated transformation of the covariant derivative $D_\mu$ must be reformulated.
By applying the covariant derivative upon a generic field $\phi$,
i. e. $D^{\prime}_\mu \phi$, with $\phi$ gauge transformation given by
\begin{equation}
\phi = U \Psi V \Rightarrow  U^{\dagger} \phi V^{\dagger} = \Psi,
\end{equation}
we construct the simplest and most succinct form of describing
the $SU(n)_{U} \otimes SU(m)_{V}$ gauge transformation of
$D_\mu$ represented by
\begin{equation}
D^{\prime}_\mu \phi = U (D_\mu U^{\dagger} \phi V^{\dagger}) V
\label{srtss}
\end{equation}
from which we cannot factorize the field $\phi$.
Since we know the covariant derivative transformation properties, we can write
\begin{equation}
D_\mu \Psi = \partial_\mu \Psi + g_{U} W^{U}_\mu \Psi + g_{V} \Psi W^{V}_\mu. 
\label{dfi}
\end{equation}
with the anti-Hermitian operators $W^{\U(\V)}_\mu$
\footnote{In order to maintain the validity of the Hermitian properties of the total Lagrangian,
the operators described by $W^{\U(\V)}_\mu$ must be completely anti-Hermitian.} defined by
\begin{equation}
W^{\U(\V)}_\mu = -i T\,^{\U(\V)}_a W_{a\mu},
\label{dfii}
\end{equation}
where $T_a^{\U(\V)}$ are the generators of the $SU(n)_{\U(\V)}$ algebra and $W_{a\mu}$ represent
the vector gauge bosons. 
From expressions (\ref{dfi}) and (\ref{srtss}), we can write
\begin{eqnarray}
\partial_\mu \phi + g_{U} W^{U\prime}_\mu \phi + g_{V} \phi W^{V\prime}_\mu &=& U (D_\mu U^{\dagger} \phi V^{\dagger}) V \nonumber \\
&=& U (\partial_\mu U^{\dagger} \phi V^{\dagger}) V + (g _{U} U W^{U}_\mu U^{\dagger} \phi V^{\dagger} V + g_{V} U U^{\dagger} \phi V^{\dagger}W^{V}_\mu V)\nonumber\\ 
&=& \partial_\mu \phi + U (\partial_\mu U^{\dagger}) \phi  + \phi(\partial_\mu V^{\dagger})V \nonumber\\
&&~~~~~~~~~~~~~~~~~~+ (g _{U} U W^{U}_\mu U^{\dagger}\phi + g_{V} \phi V^{\dagger}W^{V}_\mu V). 
\label{dfia}
\end{eqnarray}
and thus, we obtain
\begin{eqnarray}
W^{U \prime}_\mu  &=& \frac{1}{g_{U}} U (\partial_\mu U^{\dagger})  + U W^{U}_\mu U^{\dagger}\\
\label{dfib0}
W^{V \prime}_\mu &=& \frac{1}{g_{V}} (\partial_\mu V^{\dagger})V   + V^{\dagger}W^{V}_\mu V, 
\label{dfib0A}
\end{eqnarray}
so that the local gauge invariance of the fermionic kinetic term of the Lagrangian will be completely satisfied in (\ref{defdef}).

By following the above analysis, we can construct the gauge boson kinetic term 
once we have easily observed that
\begin{eqnarray}
D_\nu(D_\mu \Psi) &\Rightarrow& D_\nu^{\prime}(D_\mu^{\prime} \Psi^{\prime})\nonumber\\
				  &     =     & D_\nu^{\prime}(U (D_\mu \Psi) V) \nonumber\\
				  &		= 	  & U (D_\nu U^{\dagger} U (D_\mu \Psi)V V^{\dagger})V \nonumber\\
				  &		=	  & U (D_\nu (D_\mu \Psi)) V. 
\label{defc}
\end{eqnarray}
i. e., the term given by $D_\mu(D_\nu \Psi)$ as well as the commuting term $[D_\mu, D_\nu]\Psi$,
by simple analogy, transforms exactly as $\Psi$ does.
By following the definition of the covariant derivative,
we can write
\begin{equation}
[D_\mu, D_\nu] \phi = g_{U} F^{U} _{\mu \nu} \phi + g_{V} \phi F^{V}_{\mu\nu}. 
\label{defd}
\end{equation}
where
\begin{eqnarray}
F^{\U(\V)}_{\mu \nu} &=& \partial_\mu W_\nu^{\U(\V)} - \partial_\nu W_\mu^{\U(\V)}
+ g_{\U(\V)}[W^{\U(\V)}_\mu,W^{\U(\V)}_\nu], 
\label{defe}
\end{eqnarray}
so that we can immediately obtain the invariant $SU(n)_{\U(\V)}$ gauge boson kinetic term,
\begin{eqnarray}
\mathcal{L}_2 &=& \frac{1}{2}Tr [F^{\U(\V)\prime}_{\mu\nu}F^{\U(\V) \mu \nu \prime}]
=\frac{1}{2}Tr [F^{\U(\V)}_{\mu\nu}F^{\U(\V) \mu \nu}],
\label{deff}
\end{eqnarray}
Obviously for the kinetic term for scalar bosons $\phi_a$ which transforms as 
a $SU(n)_{\U(\V)}$ singlet the Lagrangian invariance is immediately obtained.
To summarize, the complete kinetic part of the Lagrangian invariant under the local gauge transformation of
$SU(n)_{U} \otimes SU(m)_{V}$ can thus be described by
\begin{equation}
\mathcal{L} = i\, Tr [\overline{\Psi} (\gamma^\mu D_\mu) \Psi] - 
\frac{1}{2}Tr [F^{V \dagger}_{\mu\nu}F^{V \mu \nu}] -
 \frac{1}{2}Tr [F^{U \dagger}_{\mu\nu}F^{U \mu \nu}] + Tr[(D_\mu\phi_a)^{\dagger}D_\mu \phi_a].
\label{defg}
\end{equation}

Since we have learned the gauge transformation rules for constructing the fermion and boson
invariant Lagrangian kinetic terms of a $SU(n)_{U} \otimes SU(m)_{V}$ theory, 
we can apply such a procedure for constructing 
the $SU(2)_{L} \otimes U(1)_{Y}$ Lagrangian kinetic term 
of the GWS theory for electroweak interactions, i. e.
\begin{equation}
\mathcal{L} = i \,Tr [\overline{\Psi} (\gamma^\mu D_\mu) \Psi] - \frac{1}{2}Tr [F^{\dagger}_{ \mu\nu}F^{\mu \nu}] - \frac{1}{2}Tr [F^{\B \dagger}_{\mu\nu}F^{\B \mu \nu}] + Tr[(D_\mu\phi)^{\dagger}D_\mu \phi].
\label{defh}
\end{equation}
The fermionic fields of the theory transform as
\begin{equation}
\Psi \rightarrow \Psi^{\prime} = U \Psi V,
\end{equation}
for which have explicitly defined
\begin{equation}
U = \exp[{i \, g \,\alpha_a(x) \,T_a}]
\end{equation}
and
\begin{equation}
V = \exp{[i \, g^{\prime}\, \beta(x) \,\frac{Y}{2}]}
\end{equation}
where $Y$ is the quantum number of hypercharge of the $U(1)_{Y}$ symmetry and $T_i =\frac{\sigma_i}{2}$
are the $SU(2)_{L}$ generators written in terms of Pauli matrices $\sigma_i$.
By considering that the $V$ operations of $U(1)_{Y}$ act as diagonal matrices, we can rewrite the
covariant derivative $D_\mu$ as 
\begin{equation}
D_\mu \phi = \, (\partial _\mu - i\, g W_{a \mu} T_a - i\, g^{\prime}\frac{Y}{2} B_\mu) \phi
\label{f3},
\end{equation}
and the tensorial variables $F_{\mu\nu}$ as
\begin{eqnarray}
i\, F^{\mu\nu} &=& F_i^{\mu\nu}T_i 
= \left(\partial ^\mu W_i^\nu - \partial ^\nu W_i^\mu\right)T_i
 -i \,g [T_j,T_k]\,W_j^\mu W_k^\nu
\label{ff3}
\end{eqnarray}
and
\begin{equation}
i \, F^{\B \mu\nu} = \left(\partial^\mu B^\nu - \partial^\nu B^\mu\right).
\label{ff3A}
\end{equation}
The fermionic fields 
$\Psi \equiv \Psi_{\L}$, with negative (left) chirality, transform as $SU(2)_{L}$ doublets
and 
$\Psi \equiv \Psi_{\R}$, with positive (right) chirality, transform as $SU(2)_{L}$ singlets.
The scalar field transform as an $SU(2)_{L}$ doublet and has a non-null vacuum expectation value.
The gauge boson vector fields $W^i_\mu$ which mediate the electroweak chiral interactions transform
as the $SU(2)_{L}$ vector representation $T^i$ and the boson $B^\mu$ is related
to the unique symmetry of $U(1)_{Y}$.
Under the simultaneous gauge transformations represented by
\begin{eqnarray} 
W_i^\mu T_i &\rightarrow& W^\mu_{i \prime}T_i = W^\mu_iT_i - \frac{1}{g}(\partial ^\mu \alpha_iT_i  -  i\, g[T_b,T_c] W^\mu_b \alpha_c),\nonumber\\
B^\mu       &\rightarrow& B^\mu_{  \prime}    = B^\mu      - \frac{1}{g^{\prime}}(\partial ^\mu \beta),\nonumber \\
\psi		&\rightarrow& \psi^{\prime} = \exp{\left[-i \,(\alpha_i T_i \frac{1-\gamma^5}{2}+ \beta \frac{Y}{2}) \right]}\psi, \nonumber \\
\phi		&\rightarrow& \phi^{\prime} = \exp{\left[-i \,(\alpha_i T_i + \beta \frac{Y}{2}) \right]}\phi.
\label{ffff3}
\end{eqnarray}
the Lagrangian of expression (\ref{f3}) is locally invariant.
The vacuum expectation value of the scalar field $\phi$ continue to be invariant under the gauge transformation
where $\alpha_1 = \, \alpha_2 = \, 0~$ e $~\alpha_3 = \, \beta$.
By using the parameterization
\begin{eqnarray}
\phi &=& \frac{1}{\sqrt{2}}\exp\left[\frac{2i}{\upsilon} (\xi_i T_i)\right]\left(\begin{array}{c} 0\\ \upsilon + \eta \end{array}\right)
	= \frac{1}{\sqrt{2}} \left(\begin{array}{c} \xi_2 + i\xi_1\\ \upsilon + \eta -i\xi_3 \end{array}\right),
\end{eqnarray}
we can observe that the vacuum expectation value given by
\begin{equation}
\langle\phi\rangle = \, \phi _0 = \, \frac{1}{\sqrt{2}}\left(\begin{array}{c} 0\\ \upsilon \end{array}\right)
\label{f2},   
\end{equation}
leads to a spontaneous symmetry breaking which reduces the system symmetry from $SU(2)_{L} \otimes U(1)_{Y}$ to $U(1)_{EM}$.

The GWS theory corresponds to one of the simplest cases, 
but when we treat a generic $SU(n)_{U} \otimes SU(m)_{V}$ theory, the matrix ($U$ and $V$) notation
elegantly developed in this manuscript becomes essentially more convenient, as we have tried to demonstrate.
For ones which are interested in unification theory models \cite{Moh86}, the mathematical structure over which we have elaborated on
can be useful for exploring extensions of some unification models, for instance, the
the Mohapatra-Pati $SU(2)_{L} \otimes SU(2)_{R} \otimes U(1)_{B-L}$
\cite{Moh75} or the Pati-Salam $SU(4)_{c} \otimes SU(2)_{L} \otimes SU(2)_{R}$ \cite{Pat74}.
In particular, a double-sided transformation of spinor fields has just been considered in the standard model context,
but in the particular case of $SU(3) \otimes SU(2)$ maps \cite{Tra01}.

To summarize, in this manuscript we have reported about bilateral transformations on the spinor field, related to $SU(n)_{U} \otimes SU(m)_{V}$ maps.
From a gauge viewpoint, in the context of a spin-Clifford bundle, we have presented a self-contained theoretical construction.
Such a generalization is important mainly because it can make us to better understand the transformations of spinor fields in a gauge theory, which certainly contributes to enlarge the application spectrum of the present theory.
We suppose that the above material can contribute to the study of gauge transformations applied to generalized bilateral maps as well as for complementing an introductory study of the standard model of elementary particles.

{\bf Acknowledgments}
The author would like to thank FAPESP (PD 04/13770-0) for the financial support.

\end{document}